\documentclass[aps,showpacs,prl,twocolumn]{revtex4}

\begin{document}

\preprint{{DOE/ER/40762-275}\cr{UMPP\#03-036}}


\count255=\time\divide\count255 by 60 \xdef\hourmin{\number\count255}
  \multiply\count255 by-60\advance\count255 by\time
 \xdef\hourmin{\hourmin:\ifnum\count255<10 0\fi\the\count255}

\newcommand{\xbf}[1]{\mbox{\boldmath $ #1 $}}

\newcommand{\sixj}[6]{\mbox{$\left\{ \begin{array}{ccc} {#1} & {#2} &
{#3} \\ {#4} & {#5} & {#6} \end{array} \right\}$}}

\newcommand{\threej}[6]{\mbox{$\left( \begin{array}{ccc} {#1} & {#2} &
{#3} \\ {#4} & {#5} & {#6} \end{array} \right)$}}

\title{New Relations for  Excited Baryons in Large $N_c$ QCD}

\author{Thomas D. Cohen}
\email{cohen@physics.umd.edu}

\affiliation{Department of Physics, University of Maryland, College
Park, MD 20742-4111}

\author{Richard F. Lebed}
\email{Richard.Lebed@asu.edu}

\affiliation{Department of Physics and Astronomy, Arizona State
University, Tempe, AZ 85287-1504}

\date{January, 2003}

\begin{abstract}
We show that excited baryons in large $N_c$ QCD form multiplets,
within which masses are first split at $O(1/N_c)$.  The dominant
couplings of resonances to various mesons are highly constrained: The
$N(1535)$ decays at leading $1/N_c$ order exclusively to $\eta$-$N$
rather than $\pi$-$N$, and vice versa for the $N(1650)$.  This
multiplet structure is reproduced by a simple large $N_c$ quark model, well studied in the literature, that describes resonances as
single-quark excitations.
\end{abstract}

\pacs{11.15.Pg, 12.39.-x, 13.75.Gx, 14.20.Gk}

\maketitle

During the past two decades there have been two approaches to using
large $N_c$ QCD in the description of excited baryons.  The first
approach~\cite{HEHW,MP,MM,Mat3,MK}, developed in the mid-1980s in the
context of Skyrme models, was based on the study of meson-baryon
scattering with the identification of excited baryons as resonances
(as indeed they are in nature).  The second
approach~\cite{CGKM,PY,Goity,CCGL1,CCGL2,GSS,CC1,CC2}, developed more
recently, was cast in quark model language and used operator counting
rules in a manner similar to those developed to describe the large
$N_c$ properties of the ground-state band of $I=J$
baryons~\cite{DM,Jenk,DJM1,DJM2,JL}.  However, unlike for the case of
ground-state band baryons, where the quark model is a shorthand for
simply implementing group theory, the quark model treatment requires
nontrivial dynamical assumptions.  In particular, the method is based
on matrix elements of operators between excited baryon states, which
is strictly only sensible if the excited states are stable, at least
at large $N_c$.  While this assumption is true in quark models, it is
not true in large $N_c$ QCD. This raises the question of the extent to
which results for excited baryons in such treatments are in fact
results of large $N_c$ QCD.

Here we show: i) At leading order in the large $N_c$ expansion of QCD
the baryons form multiplets, within which masses and widths are
degenerate up to splittings of $O(1/N_c)$; this result can be derived
from large $N_c$ consistency rules {\it with no additional model
assumptions}.  ii) Certain multiplets have no coupling to the
$\pi$-$N$ decay channel, while others have no coupling to the
$\eta$-$N$ channel up to corrections of $O(1/N_c)$; phenomenological
evidence of this is seen in the decays of the $N(1535)$ and $N(1650)$.
The degeneracy structure of result i) was previously derived in the
quark model language by Pirjol and Yan~\cite{PY}. We reproduce these
quark model results via the technology of Refs.~\cite{CCGL1,CCGL2}. We
note that the fact that large $N_c$ quark model reproduces the
multiplet structure of large $N_c$ QCD helps to justify these models.

In this work we focus on the lowest-lying negative-parity states.  In
the $N_c=3$ quark model they correspond to a {\bf 70}-plet of SU(6).
For simplicity we look here at nonstrange baryons.  However, the
generalization to positive-parity states and states with nonzero
strangeness is straightforward and will be considered in the
future~\cite{future}.

We start by working directly in a large $N_c$ world (neglecting
$1/N_c$ corrections) in order for our analysis to be consistent with
results based on treatments of meson-baryon scattering.  Of course, in
comparing with the physical world of $N_c=3$, one must bear in mind
possible $1/N_c$ corrections (which can in principle be parameterized
in a consistent and constrained way), as well as artifacts such as
states existing in large $N_c$ but not in the physical world.

Our principal tools are the linear relations between meson-baryon
scattering amplitudes in various partial
waves~\cite{HEHW,MP,MM,Mat3,MK} that become exact in the large $N_c$
limit.  Here we limit our attention to the scattering of $\pi$ and
$\eta$ mesons off ground-state band ($I=J$) baryons, in which the
final state contains the same meson as in the initial state.  Recall
that in the large $N_c$ limit the $\Delta$ is stable, and thus it is
perfectly meaningful to discuss scattering off the $\Delta$.  The
relevant $S$ matrix formulas~\cite{HEHW,MP,MM,Mat3,MK} for such
scattering are given by:
\begin{eqnarray}
S_{LL^\prime R R^\prime IJ}^\pi & \! = &\sum_K (-1)^{R^\prime \! \!
- R} \sqrt{(2R+1)(2R^\prime+1)} (2K+1) \nonumber \\
& & \times \left\{ \begin{array}{ccc} K &
I & J\\ R^\prime & L^\prime & 1 \end{array} \right\} \left\{
\begin{array}{ccc} K & I & J \\ R & L & 1 \end{array} \right\}
s_{K L^\prime L}^\pi , \label{MPeqn1} \\
S_{L R J}^\eta & = & \sum_K
\delta_{KL} \, \delta (L R J) \, s_{K}^\eta .\label{MPeqn2}
\end{eqnarray}
For $\pi$ scattering, we denote the incoming baryon spin (=isospin) by
$R$, the final baryon spin as $R^\prime$, the orbital angular momentum
of the incident (final) pion by $L$ ($L^\prime$), and $I$ ($J$)
represents the total isospin (angular momentum) of the state.
$S_{LL^\prime R R^\prime IJ}^\pi$ denotes the $S$ matrix for this
channel (isospin- and angular momentum-reduced in the Wigner-Eckart
theorem sense), the factors in braces are $6j$ coefficients, and $s_{K
L^\prime L}^\pi$ are universal amplitudes that are independent of $I$,
$J$, $R$, and $R'$.  In $\eta$-meson scattering, the isospin $R$ of
the baryon cannot change and equals the total isospin $I$ of the
state.  The orbital angular momentum $L$ of the $\eta$ remains
unchanged in the process due to large $N_c$ constraints.  The total
angular momentum $J$ of the state is then constrained by the triangle
rule $\delta (L R J)$.  $S_{L R J}^\eta$ is the reduced scattering
amplitude, and $s_{K}^\eta$ are universal amplitudes independent of
$R$ and $J$.  The structure of Eqs.~(\ref{MPeqn1}) and (\ref{MPeqn2})
imply that the scattering amplitudes in different channels are
linearly related.  There are more $S$ matrix amplitudes $S_{LL^\prime
R R^\prime IJ}^\pi$ than there are $s_{K L^\prime L}^\pi$ functions;
thus there are linear constraints between the $S_{LL^\prime R R^\prime
IJ}^\pi$ (that hold to leading order in the $1/N_c$
expansion). Analogously, there are fewer $s_{K}^\eta$ amplitudes than
$S_{L R J}^\eta$ amplitudes.

Equations~(\ref{MPeqn1}) and (\ref{MPeqn2}) were originally
derived~\cite{HEHW,MP,MK,MM,Mat3} in the Skyrme model. 
Despite these initial derivations, the validity of Eqs.~(\ref{MPeqn1})
and (\ref{MPeqn2}) does not depend on model assumptions about chiral
solitons; rather the equations follow directly from large $N_c$ QCD.
A derivation of these relations directly from the large $N_c$
consistency rules, based on the formalism of Ref.~\cite{DJM2} and
which exploits the noted $I_t=J_t$ rule~\cite{MM}, can be found in a
more detailed report of our work~\cite{article}. We note that the only
assumptions needed in this derivation are i) Baryonic quantities in
large $N_c$ QCD scale according to the generic large $N_c$ rules of
Witten~\cite{Wit}, or more slowly (if there are cancellations); ii)
there exists a hadronic description that reproduces the large $N_c$
QCD results; iii) the $\pi$-$N$ coupling scaling is generic (without
cancellations) scaling as $N_c^{1/2}$; and iv) nature is realized in
the most symmetric representation of the contracted $SU(2 N_f)$ group
that emerges from the previous assumptions.

To make concrete predictions about degeneracy patterns, one needs a
method to identify the position of a resonance. We use a common
theoretical definition of resonance position: The partial-wave
scattering amplitudes are functions of energy.  By analytically
continuing the energies into the complex plane one can identify the
resonance position as the location of the pole, with the real part
being identified as the mass and the imaginary part as the width.
This theoretical definition is unambiguous but has the disadvantage of
not being directly determined by scattering with physical kinematics.
Fortunately, in the limit of vanishingly small widths, the theoretical
definition agrees with the result extracted from scattering data via
any sensible prescription.

This resonance position definition implies the existence of
degeneracies. Consider the structure of Eqs.~(\ref{MPeqn1}) and
(\ref{MPeqn2}).  A pole in the partial wave amplitude for some channel
on the left-hand side of Eq.~(\ref{MPeqn1}) implies a pole in one of
the $s^\pi_{K L^\prime L}$ amplitudes on the right-hand side.  But
since all the $s^\pi_{K L^\prime L}$ amplitudes contribute to more
than one channel, a pole in one channel implies the existence of poles
in others, {\it i.e.}, states with different $I,J$ quantum numbers but
with the same mass and width.  An analogous argument exists for
$\eta$-meson scattering.

To isolate multiplets of degenerate baryon states, one must first
determine whether degeneracies could occur in poles of the $s^\pi_{K
L^\prime L}$ and $s^\eta_K$ amplitudes themselves.  {\it A priori\/}
one expects the $K$ sectors ought to be dynamically disconnected at
large $N_c$, as it would be unnatural to suppose otherwise.  In Skyrme
models, this is quite clear---the various $K$ sectors are completely
separate.  However, one may reasonably expect degeneracies in the
poles of amplitudes with identical $K$ but distinct values of $L$ or
$L^\prime$: $L$ is not a conserved quantum number; one can have a
resonance with given $I$ and $J \,$ but that can reached in scattering
by more than one $L$ ({\it e.g.}, by scattering off a $\Delta$ rather
than $N$).  In such a case, scattering amplitudes in channels with
different $L$ have degenerate poles.  Similarly, partial-wave
amplitudes involving the scattering of different mesons, but which
couple to the same $K$ channel, may produce degenerate poles ({\it
e.g.}, $s^\pi_{222}$ and $s^\eta_2$).

Degenerate multiplets can be found quite simply.  Start by assuming a
pole in one of the $s^\pi_{K L^\prime L}$ or $s^\eta_K$ amplitudes and
use selection rules implicit in the $6j$ coefficients in
Eq.~(\ref{MPeqn1}) and the triangle constraint in Eq.~(\ref{MPeqn2})
to find all partial-wave amplitudes that couple.  Note in doing this
one must consider all possible $L$, $L^\prime$, $R$, and $R^\prime$ as
well as the conserved $I$ and $J$ (which label the quantum numbers of
the state).  Since we focus on negative-parity states, we restrict
attention to the case of $L$ even. From this exercise one identifies
the following multiplets of negative-parity baryon states with
degenerate masses and widths (modulo splitting at order $1/N_c$) and
the associated $s^\pi_{K L^\prime L}$ or $s^\eta_K$ amplitudes:
\begin{eqnarray}
N_{1/2} , \, \Delta_{3/2} , &\!\!\cdots\!\!&
(s_{0}^\eta) , \label{s0}\\
N_{1/2} , \, \Delta_{1/2} , \, N_{3/2} ,
\Delta_{3/2} , \, \Delta_{5/2} , &\!\!\cdots\!\!&
(s_{1 0 0}^\pi, s_{1 2 2}^\pi) , \hspace{1.5em} \label{s1}\\
\Delta_{1/2} , \, N_{3/2} , \, \Delta_{3/2} ,
\, N_{5/2} , \, \Delta_{5/2} , \, \Delta_{7/2} , &\!\!\cdots\!\!&
(s_{2 2 2}^\pi, s_{2}^\eta) , \label{s2} \\
\Delta_{3/2} , \, N_{5/2} , \, \Delta_{5/2} , \, \Delta_{7/2} ,
&\!\!\cdots\!\!& (s_{3 2 2}^\pi) . \label{s3}
\end{eqnarray}
These multiplets are formally infinite in size, as we are working in
the large $N_c$ limit and thus some states are large $N_c$ artifacts.
Note that all amplitudes with the same $K$ have exactly the same
multiplet structure: For example, $s^\pi_{1 0 0}$ and $s^\pi_{1 2 2}$
have the same states in their multiplets, as do $s_{2 2 2}^\pi$,
$s_{2}^\eta $.  As discussed above, it is natural to interpret this to
mean that any given multiplet of baryon states is accessible via more
than one scattering channel provided it has the same $K$.

We note that the same degeneracy patterns were found using the quark
model assumption of stable excited baryons~\cite{PY}.  However, as
noted above this assumption is inconsistent with large $N_c$ QCD where
the width of excited states in generically $O(N_c^0)$.  We stress that
our prediction of these degenerate multiplets is model independent,
following directly from large $N_c$ consistency conditions and that it
predicts degenerate masses and widths. We do not predict {\it a
priori\/} whether or not large $N_c$ QCD generates low-lying
resonances that are narrow enough to identify.  Rather we show that if
any such resonances do exist, they fall into multiplets labeled by
$K$. The fact that we find multiplets of degenerate states at large
$N_c$ is not surprising; it reflects the contracted $SU(2N_f)$
symmetry emerging for baryons as $N_c \rightarrow \infty$.  It is
somewhat surprising that these multiplets were not identified
previously in the context of large $N_c$ meson-baryon scattering
(excepting the degeneracy between the $N_{1/2}$ and the
$\Delta_{1/2}$), particularly since studies of the baryon spectrum in
models using Eq.~(\ref{MPeqn1}) were done nearly two decades
ago~\cite{HEHW,MK}.  In those works, however, the resonance position
was fixed ``experimentalist-style'' via motion in the Argand plots
rather than by looking directly at poles in the complex plane; this
obscured the underlying degeneracies.

We note that Eqs.~(\ref{s0})--(\ref{s3}) may be of more theoretical
interest than phenomenological import in the real world of $N_c$=3.
The phenomenological difficulty is that the states in nature one may
wish to identify with the negative-parity ${\bf 70}$-plet (in quark
model language) lie in a 200 MeV-wide band (from 1520 MeV ($N_{3/2}$)
to 1700 MeV ($\Delta_{3/2}$), as listed by the Particle Data
Group~\cite{PDG}).  Such states may be associated with three distinct
$K$'s, all of which are split by $O(N_c^0)$.  On the other hand, the
$\Delta$-$N$ mass splitting is an $O(1/N_c)$ effect but is $\sim
290$~MeV.  Since the expected $1/N_c$ corrections that split the
multiplets are of the same scale or larger than the distance between
the multiplets, it may be very difficult to identify states clearly as
belonging to given $K$ multiplets on the basis of spectroscopy alone.

However, the multiplet structure does have nontrivial phenomenological
implications for the {\it decays\/} of excited baryons. Note that
$K=0$ negative-parity states do not couple to the $\pi$-$N$ channel.
This follows from the fact [Eq.~(\ref{MPeqn1})] that $K$ is a vector
sum of $L$ and the $\pi$ isospin (=1), implying that $K=0$ only occurs
for $L=1$, which gives the wrong parity. Similarly, the
negative-parity states in the $K=1$ multiplet cannot couple to the
$\eta$-N channel.  Assuming that there exist well isolated resonances
in both the $K=0$ and $K=1$ channels, at large $N_c$ the states
associated with the $K=1$ channel decay into $\pi$-$N$ but not
$\eta$-$N$, and vice versa for the $K=0$ states.  For the $N_c=3$
world there are subleading $1/N_c$ effects, which can cause some
mixing of the $K$ modes and hence give nonzero but suppressed decays
into the ``forbidden'' channels.  When looking for this effect in
phenomenology one must avoid large $N_c$ artifacts.  In particular,
the existence of more than one low-lying negative-parity
$\Delta_{3/2}$ state appears to be an artifact (only one is present in
the $N_c=3$ quark model), so this prediction for decay modes only
applies to the $N_{1/2}$ in the real world.  The decay patterns of the
$N(1535)$ and $N(1650)$ are consistent with the $N(1535)$ being
predominantly the $K=0$ state, while the $N(1650)$ is predominantly
$K=1$.  The Particle Data Group~\cite{PDG} lists the decay fraction of
$N(1535)$ into $\pi$-$N$ (35--55\%) as being essentially the same as
the decay fraction into $\eta$-$N$ (35--50\%), despite having a
substantially larger phase space for decay (nominally by a factor of
$\sim 2.6$ evaluated at the mass peak, but effectively higher since
the $\eta$-$N$ channel is only 50 MeV from threshold).  This indicates
the coupling to $\eta$-$N$ is much larger than to $\pi$-$N$.  In
contrast, the decay fractions for the $N$(1650) are listed
as~\cite{PDG} 55--90\% into $\pi$-$N$ and only 3--10\% into
$\eta$-$N$.  Since the phase space for decay is only $\sim 1.6$ times
greater for the $\pi$-$N$ channel, this indicates a much stronger
coupling of the state to $\pi$-$N$.

We now turn to the other major approach that has been used to study
excited baryons in the large $N_c$
limit~\cite{CGKM,PY,Goity,CCGL1,CCGL2,GSS,CC1,CC2}.  This approach
uses quark model language and expresses the large $N_c$ constraints
via identification of the $N_c$ scaling of operators.  Formally, this
is very similar to the techniques of Ref.~\cite{DJM2}.  An important
distinction, however, is that Ref.~\cite{DJM2} used quark model
language only as simple tool for doing group theory and without any
dynamical assumptions of the quark model, while the work of
Refs.~\cite{CGKM,PY,Goity,CCGL1,CCGL2,GSS,CC1,CC2} depends in part on
quark model dynamics: it neglects coupling to decay channels and hence
the widths of the baryons.  Moreover, the analyses done so far are
based on the dynamics of the simplest version of the quark model, {\it
i.e.}, the lowest-lying negative-parity states are described as being
single-quark excitations without configuration mixing. In this picture
the states discussed here have one quark carrying a single unit
($\ell=1$) of orbital excitation about an $(N_c\!-\!1)$-quark core
symmetric under spin$\times$flavor.  For $N_c=3$ this produces the
familiar SU(6) {\bf 70}-plet, but for larger $N_c$ generates
additional states.

Previous derivations of the multiplets were based directly on
implementation of the large $N_c$ consistency rules~\cite{PY} which,
though intellectually elegant, is rather cumbersome.  Here we show how
to derive them us the simple operator methods of
Refs.~\cite{CCGL1,CCGL2}.  Up to $O(N_c^0)$, 3 operators contribute to
the Hamiltonian in this quark picture, denoted by ${\cal H} = c_1
\openone + c_2 \ell s + c_3 \ell^{(2)} g G_c/N_c$, where lowercase
indicates operators acting upon the excited quark, subscript $c$
indicates those acting upon the core, $G^{ia}$ denotes the combined
spin-flavor operator $\propto q^\dagger \sigma^i \tau^a q$, and
$\ell^{(2)}$ is the $\Delta \ell = 2$ tensor
operator. References~\cite{CCGL1,CCGL2} elucidate this notation.  We
find that to leading order in $1/N_c$ all of the mass eigenvalues are
given by only three linear combinations of these parameters:
\begin{eqnarray}
m_0 & \equiv & c_1 N_c - ( c_2 + 5/24 \, c_3 ), \nonumber \\
m_1 & \equiv & c_1 N_c - 1/2 \, ( c_2 - 5/24 \, c_3 ), \nonumber \\
m_2 & \equiv & c_1 N_c + 1/2 \, ( c_2 - 1/24 \, c_3 ) .
\end{eqnarray}

For example, the Hamiltonian up to $O(N_c^0)$ for the two $N_{1/2}$
states is
\begin{eqnarray}
H_{N_{1/2}} &= &(\overline{N}_{1/2} \, \overline{N}^{\,
\prime}_{1/2}) \, {\bf M}_{N_{1/2}} ( N_{1/2} \, N^\prime_{1/2}
)^T , \nonumber \\
 {\bf M}_{N_{1/2}} &= &\left(
\begin{array}{ccc} c_1 N_c -\frac 2 3 c_2 &&
-\frac{1}{3\sqrt{2}} c_2 -\frac{5}{24\sqrt{2}} c_3 \\
-\frac{1}{3\sqrt{2}} c_2 -\frac{5}{24\sqrt{2}} c_3 && c_1 N_c
-\frac 5 6 c_2 -\frac{5}{48} c_3 \end{array} \right). \nonumber \\ &&
\label{N1mass}
\end{eqnarray}
where $N_{1/2}$ and $N^\prime_{1/2}$ refer to unmixed negative-parity
spin-1/2 nucleon states in the initial quark model basis.  This may be
obtained from Eqs.~(A6)--(A8) or Table~II of Ref.~\cite{CCGL2} [again
including only contributions up to $O(N_c^0)$].  Diagonalizing, one
finds the masses of the two physical states are given by,
$M_{N_{1/2}}^{(1)} = m_0$ and $M_{N_{1/2}}^{(2)} = m_1$.  Note the
surprising absence of square roots of terms quadratic in $c_2$ and
$c_3$, which also implies simple analytic results for mixing angles as
will be presented in our longer article on this work~\cite{article}.

Using analogous notation for the other states, we find
$M_{N_{3/2}}^{(1)} = m_2$, $M_{N_{3/2}}^{(2)} = m_1$.  The $N_{5/2}$
state is unmixed but also has a degenerate eigenvalue: $M_{N_{5/2}} =
m_2$.  For large $N_c$, the relevant $\Delta$ states of low spin are
two $\Delta_{1/2}$'s (masses $m_1$ and $m_2$), three $\Delta_{3/2}$'s
($m_0$, $m_1$, and $m_2$), two $\Delta_{5/2}$'s ($m_0$ and $m_1$), and
one $\Delta_{7/2}$ ($m_2$) (In comparison, the only $\Delta$'s for
$N_c=3$ are a single $\Delta_{1/2}$ and $\Delta_{3/2}$). These simple
analytic expressions for the masses were not noted in previous work
using this approach; those studies included the $O(1/N_c)$ terms in
the Hamiltonian and then diagonalized numerically, obscuring the
leading $O(N_c^0)$ result.

Clearly, the fact that all of the masses described by the model are
given by either $m_0$, $m_1$, or $m_2$ to leading order in the $1/N_c$
expansion implies that at large $N_c$ the various states fall into
degenerate multiplets.  These multiplets, labeled by the mass, are
given by
\begin{eqnarray}
N_{1/2} , \, \Delta_{3/2} , &\!\!\cdots\!\!& (m_0) , \label{m0}\\
N_{1/2} , \, \Delta_{1/2} , \, N_{3/2} , \, \Delta_{3/2} , \,
\Delta_{5/2} , &\!\!\cdots\!\!& (m_1) , \label{m1}\\
\Delta_{1/2} , \, N_{3/2} , \, \Delta_{3/2}, \, N_{5/2} , \,
\Delta_{5/2} , \, \Delta_{7/2} , &\!\!\cdots\!\!& (m_2 ) . \label{m2}
\end{eqnarray}
This multiplet structure is striking. Comparing the multiplet
structures in Eqs.~(\ref{m0})-(\ref{m2}) predicted for the simple
quark model with single-quark excitations to the structure seen for
resonances derived directly from large $N_c$ QCD in
Eqs.~(\ref{s0})-(\ref{s3}), it is apparent that the two are identical.
The interpretation is simply that $m_0$ represents a pole mass
appearing in the $K=0$ scattering amplitude, $m_1$ in $K=1$, and $m_2$
in $K=2$. The fact that these two pictures have the same multiplet
structure at large $N_c$ implies that that even this simplest version
of the quark model manages to capture at least some nontrivial aspects
of QCD dynamics.  This might help explain, in part, the surprising
phenomenological successes of simple quark models.

In summary, we have shown, in a fully model-independent way, that in
large $N_c$ QCD the masses and widths of excited baryons (as measured
by their pole positions) form multiplets of states with degenerate
masses and widths labeled by $K$.  The quantum numbers of states in
these multiplets are given in Eqs.~(\ref{m0})--(\ref{m2}). The $K=0$
($K=1$) multiplet decouples from the $\pi$-$N$ ($\eta$-$N$) channel as
$N_c \rightarrow \infty$, suggesting that $N(1535)$ is associated with
$K=0$ while $N(1650)$ is associated with $K=1$.  Simple quark models
describing the lowest-lying excited baryons as single-quark
excitations reproduce the same pattern of degenerate multiplets in the
large $N_c$ limit as that seen directly in large $N_c$ QCD.

The authors thank Dan Pirjol for his communication with them pointing
out the significance of Ref.~\cite{PY}.  T.D.C.\ was supported by
D.O.E.\ through grant DE-FGO2-93ER-40762; R.F.L.\ was supported by
N.S.F.\ through grant PHY-0140362.

\end{document}